\documentclass[twocolumn,pre,showpacs,superscriptaddress]{revtex4}

\usepackage{graphicx}
\usepackage{amsmath}

\begin{document}

\title{Structural and elastic properties of a confined 2D colloidal solid: a molecular dynamics study}

\author{ M. Ebrahim Foulaadvand$^{1,2}$ and  Neda Ojaghlou $^{1}$ \\
foolad@iasbs.ac.ir\\
{\small $^1$ Department of Physics, University of Zanjan, P. O. Box 45196-313, Zanjan, Iran } \\
{\small $^2$ School of Nano-science, Institute for Research in Fundamental Sciences (IPM) , P.O. Box 19395-5531, Teheran, Iran} \\
 }
\date{\today}

\begin{abstract}

We implement molecular dynamics simulations in canonical ensemble
to study the effect of confinement on a $2d$ crystal of point
particles interacting with an inverse power law potential
proportional to $r^{-12}$ in a narrow channel. This system can
describe colloidal particles at the air-water interface. It is
shown that the system characteristics depend sensitively on the
boundary conditions at the two {\it walls} providing the
confinement. The walls exert perpendicular forces on their
adjacent particles. The potential between walls and particles
varies as the inverse power of ten. Structural quantities such as
density profile, structure factor and orientational order
parameter are computed. It is shown that orientational order
persists near the walls even at temperatures where the system in
the bulk is in fluid state. The dependence of elastic
constants, stress tensor elements, shear and bulk modulii on
density as well as the channel width is discussed. Moreover, the
effect of channel incommensurability with the triangular
lattice structure is discussed. It is shown that
incommensurability notably affects the system properties. We
compare our findings to those obtained by Monte Carlo simulations
and also to the case with the periodic boundary condition along the channel width.
.

\end{abstract}
\maketitle


\section{introduction}

Colloidal crystals are a valuable model system, since the
effective interactions between colloidal particles can be
manipulated to a large extent. Furthermore, convenient techniques
to observe the structure and dynamics of such systems are
available \cite{poon,lowen,likos,pusey,zhan03}. Colloidal
dispersions under geometric confinement can serve us to understand
the effects of confinement on the ordering of various types of
nanoparticles. Related phenomena occur in a wide variety of
systems, e.g.; electrons at the surface of liquid helium that is
confined in a quasi-one-dimensional channel \cite{glasson}, dusty
plasmas \cite{lai}; hard disks \cite{pieranski,sengupta04} and
magnetorheological \cite{haghgooie} colloids under confinement
which are of great interest for various microfluidic and other
applications. Two-dimensional colloidal dispersions have been
used successfully in studies on melting in two dimensions during
the last decades \cite{strandburg,sengupta00,binder02}. In
previous studies, much attention has been paid to the generic
effect of confinement on crystalline order in $d=2$ and to the
extent and range over which the confining boundaries disturb (or
enhance, respectively) the degree of order. The effect of
external walls on phase behavior has been studied for a long time
\cite{binder74,binder83}. The confining wall can cause structural
transition such as layering transition
\cite{degennes,gau,chaudhuri05,chaudhuri08}. Another interesting
aspect of confinement is related to formation of extended defects,
solitonic staircase and standing strain wave superstructures
\cite{blaaderen,zahn00,piacente,snook10}. In this paper we intend
to gain a more insight and shed more lights onto a previously
studied problem which is a two dimensional confined colloidal
system between two walls which exert forces on the particles
\cite{binder07}. We implement molecular dynamics simulation and
compare our findings to those obtained earlier by Monte Carlo
simulations \cite{binder07}.

\section{ Description of the Problem }

Consider a $2d$ system of zero size soft disks, i.e.; point
particles, interacting under a purely repulsive force with the
inverse power law potential $U(r)= \epsilon(\frac{\sigma}{r})^p$
where $r$ denotes the distance between particles. The motivation
for taking the spatial dimension $d=2$ comes from experimental
fact that some colloidal particles with super paramagnetic cores
in the interface of water-air thin film can be described by a $2d$
system of particles interacting with the above repulsive
potential with $p=3$ \cite{zahn97,zahn00}. However, the exponent
$p$ is taken to be $12$ for computational convenience in our
paper. We recall that taking $p=3$ makes the potential long range
which is computationally inconvenient and needs special
treatment. Choosing $p=12$ has the merit that we can compare our
findings with the bulk results obtained by extensive simulations
\cite{bagchi}. We have chosen the cutoff distance $r_c=3 \sigma $
and has adopted a reduced system of units in which $\epsilon$ and
$\sigma$ are taken as unity ($k_B=1$). Now we discuss how to
represent the effect of confining walls. One choice is to take a
smooth repulsive wall located at $x=x_{wall}$, described by a wall
potential \cite{nielaba04}
$U_{wall}=\epsilon_{wall}(\frac{\sigma}{|x-x_{wall}|})^{10}$. The
motivation for a decay with the 10th power is the idea that such
a potential would result if we have a semi infinite crystal with a
power law interaction given by the above equation, but no cutoff,
and the total potential is summed over the half space
\cite{nielaba04}. We initially set the particles on the sites of
a triangular lattice which is confined between a two dimensional
channel. The channel walls are taken to be along the $y$
direction having a distance $D$ from each other. The system
length along the $y$ direction is $L$ and periodic boundary
condition is applied in the $y$ direction. The particles number
is shown by $N$ and the number density is given by
$\rho=\frac{N}{A}$ in which $A=DL$ is the channel area. Let $a_0$
denotes the lattice constant in the triangular lattice (distance
between nearest neighbours). The relation between $\rho$ and
$a_0$ is given by $\rho^{-1}=\frac{\sqrt{3}a_0^2}{2}$. Figure (1)
illustrates the choice of the geometry: the left wall lies at
$x=0$ and the right wall is located at $x=D$.

\begin{figure}[h]
\centering
\includegraphics[width=7.5cm,height=5.5cm,angle=0]{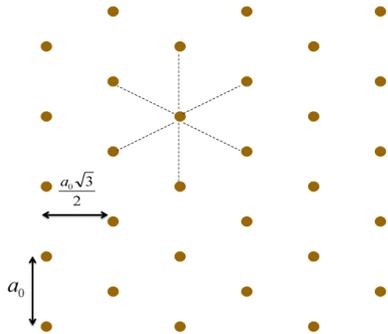}
\caption{Geometry of the problem. A 2D colloidal solid with trianagular lattice structure
is confined between two walls. The particles exerts repulsive
forces between each other. Lattice spacing is $a_0$.}\label{Fig1}
\end{figure}

We remark that one can place a triangular configuration of
particles between confining walls in two different methods. In the
first method, two of the six nearest neighbours of each particle
are located northward and southward of it whereas in the second
method two of the six nearest neighbours are located westwards and
eastwards. These two configurations are mapped into each other by
a ninety degree rotation (see figure two). As we shall see in the
rest of the paper, the elastic properties of the confined
colloidal solid differs notably for these configurations. We show
the distance between the first (last) column of particles from the
left (right) wall by $d_{L} (d_R)$ respectively. The number of
columns (rows) are denoted by $N_c$ and $N_r$ correspondingly.
Note the number of particles is given by $N=N_cN_r$. In our
simulations, we mainly have chosen
$D=(N_c+1)\frac{\sqrt{3}a_0}{2}$ with
$d_L=d_R=\frac{\sqrt{3}a_0}{2}$ but we have also studied the
incommensurate case where $D\neq (N_c+1)\frac{\sqrt{3}a_0}{2}$ (
$d_L \neq d_R$). In our simulations we have chosen
$\epsilon_{wall}=0.0005$ unless otherwise stated. It has been
shown that by this choice, the distance between columns coincide,
within error, with the ideal value $\frac{\sqrt{3}}{2}a_0$. The
readers can refer to \cite{binder07} for the further details.

\begin{figure}[h]
\centering
\includegraphics[width=7.5cm,height=5.5cm,angle=0]{fig2A.eps}
\includegraphics[width=7.5cm,height=5.5cm,angle=0]{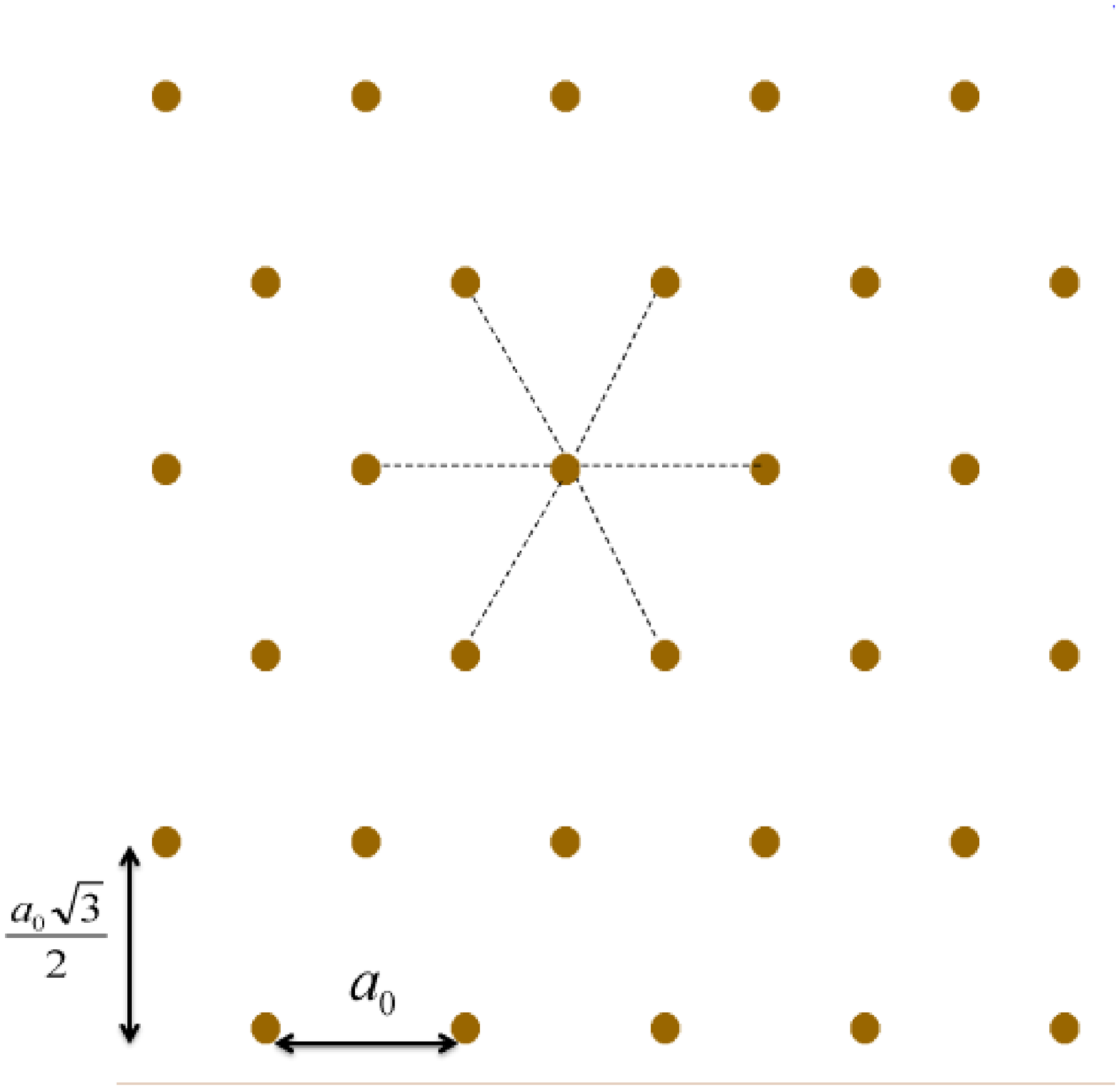}
\caption{Two methods of placing particles between the confining
walls. In method one, two of the six nearest neighbours are
located upward and downwards to the central particle (main
figure). In method two (which is rotated by ninety degree) two of
the six nearest neighbours are located leftwards and rightwards
to the central particle (upper right corner figure).}\label{Fig2}
\end{figure}

\subsection{Simulation method and details}

We have employed molecular dynamics in the NEV ensemble to
simulate the model in the reduced units. We remark that the
initial velocities are such chosen to give rise to the desired
temperature when the system reaches to a steady state. The
simulation parameters and details are as follows. The velocity
Verlet algorithm has been used for integrating the equations of
motion with a time step of $\Delta t=0.01$, the number of simulation
timestep has been mainly chosen $T=10^6$ where $2\times 10^5$
time steps are discarted for equilibration. The cut off radius
$r_c=3 \sigma$ and the shifted-force potential has been taken into
account.

\section{Structural properties}

In this section we present our simulation results for a narrow
channel. Figure (3) shows the profile of density at two
temperatures with soft wall boundary condition with $N_c=20$ and $N_r=120$ for
the method one of initial triangular setting of particles. For comparison the result
for a system with periodic boundary condition (PBC) along the $x$ direction is
shown as well. The temperature $k_BT=1$ is below the melting
point for both soft walls and PBC whereas at $k_BT=3$ the PBC
system seems to be melted but the soft wall system is not melted
yet. You see in the soft wall system, the presence of confining walls
enhances the density profile near the walls. Similar phenomenon
is observed in the Monte Carlo simulation of the problem
\cite{binder07}.

\begin{figure}[h]
\centering
\includegraphics[width=7.5cm,height=5.5cm,angle=0]{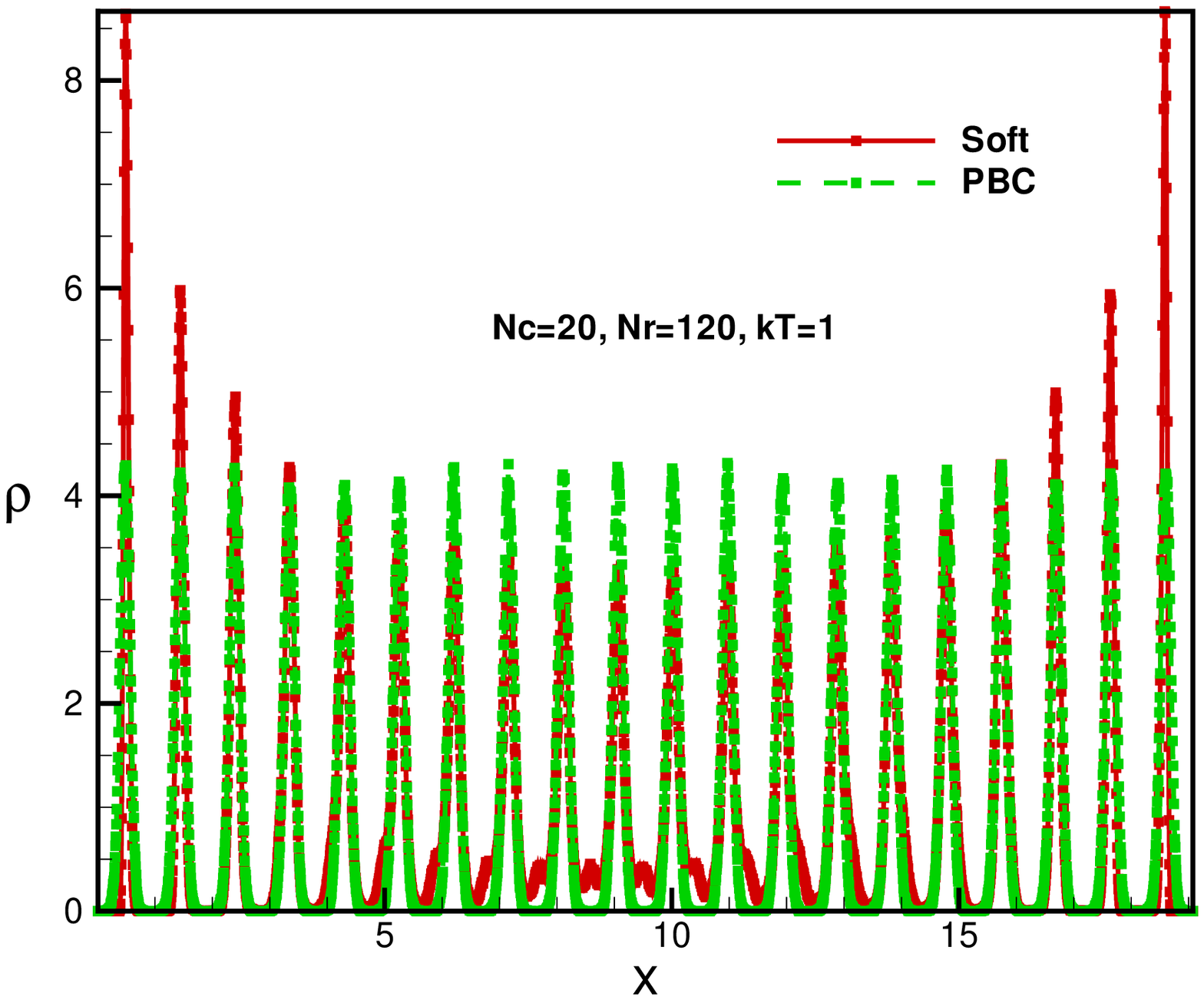}
\includegraphics[width=7.5cm,height=5.5cm,angle=0]{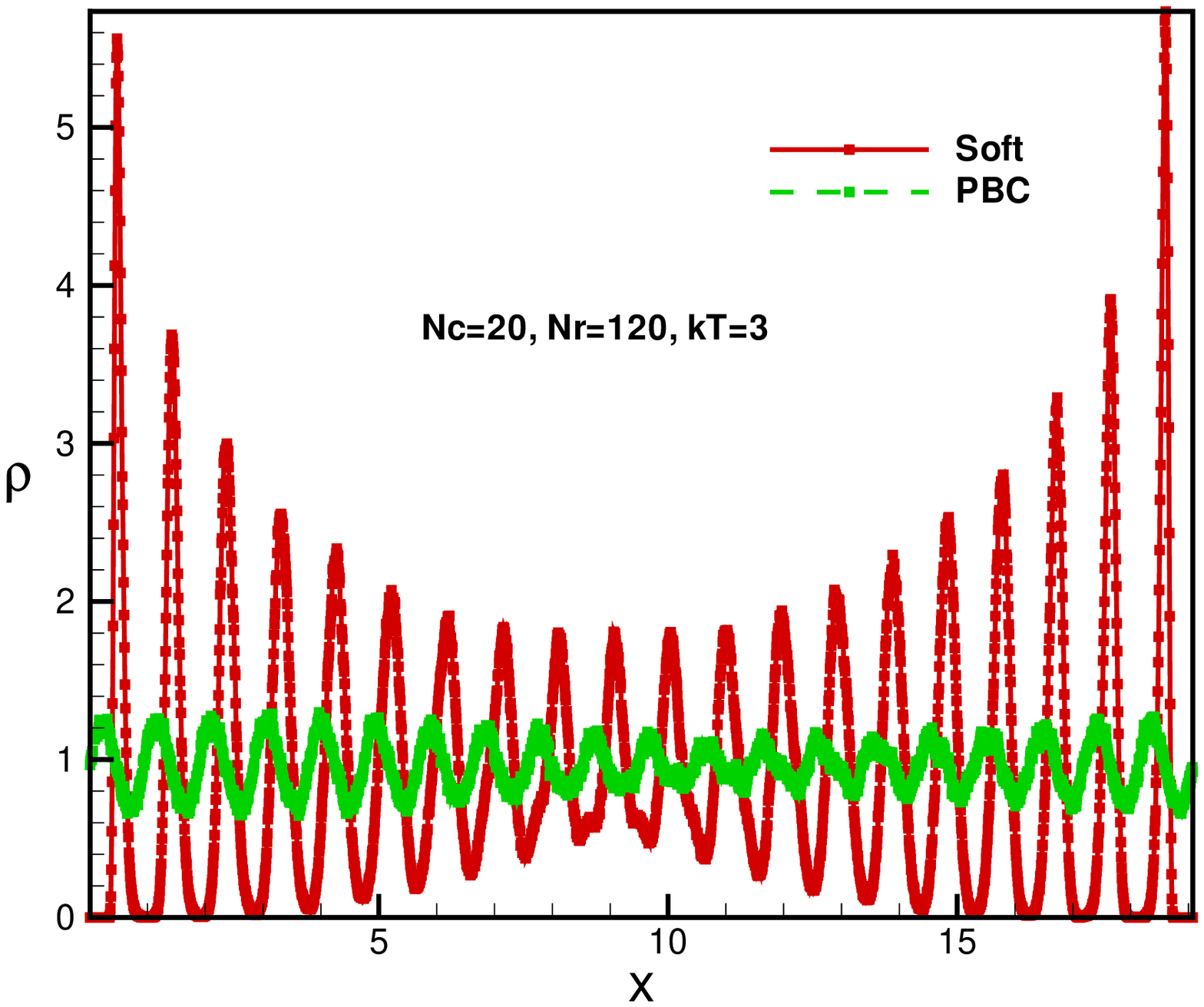}
\caption{Density profile for a narrow channel with $N_c=20$ and
$N_r=120$ at two different temperatures. Both boundary conditions
soft wall (methods one) and periodic are sketched. Top: $k_BT=1$, bottom: $k_BT=3$.}\label{Fig2}
\end{figure}

In figure (4) we compare the density profiles for methods one and two for the same temperatures as in figure (3).
At low temperatures the results are close to each other and there is no qualitative difference. When the temperature
is raised to $k_BT=3$ the difference between two methods becomes noticeable. Near walls the density profile is the same but when we leave the walls and approach the centre, the method two system melts easier than the method one system. This suggest that method one system is more stiff and exhibits a higher persistence to melting rather than method two system. The reason is due the number of particles per unit length in the adjacent column to the walls. In method one this number is proportional to $\frac{1}{a_0}$ whereas in method two this number is proportional to $\frac{1}{\sqrt{3}a_0}$ which is smaller. Consequently in method two the force per unit length exerted by a wall to its adjacent column of particles is smaller.

\begin{figure}[h]
\centering
\includegraphics[width=7.5cm]{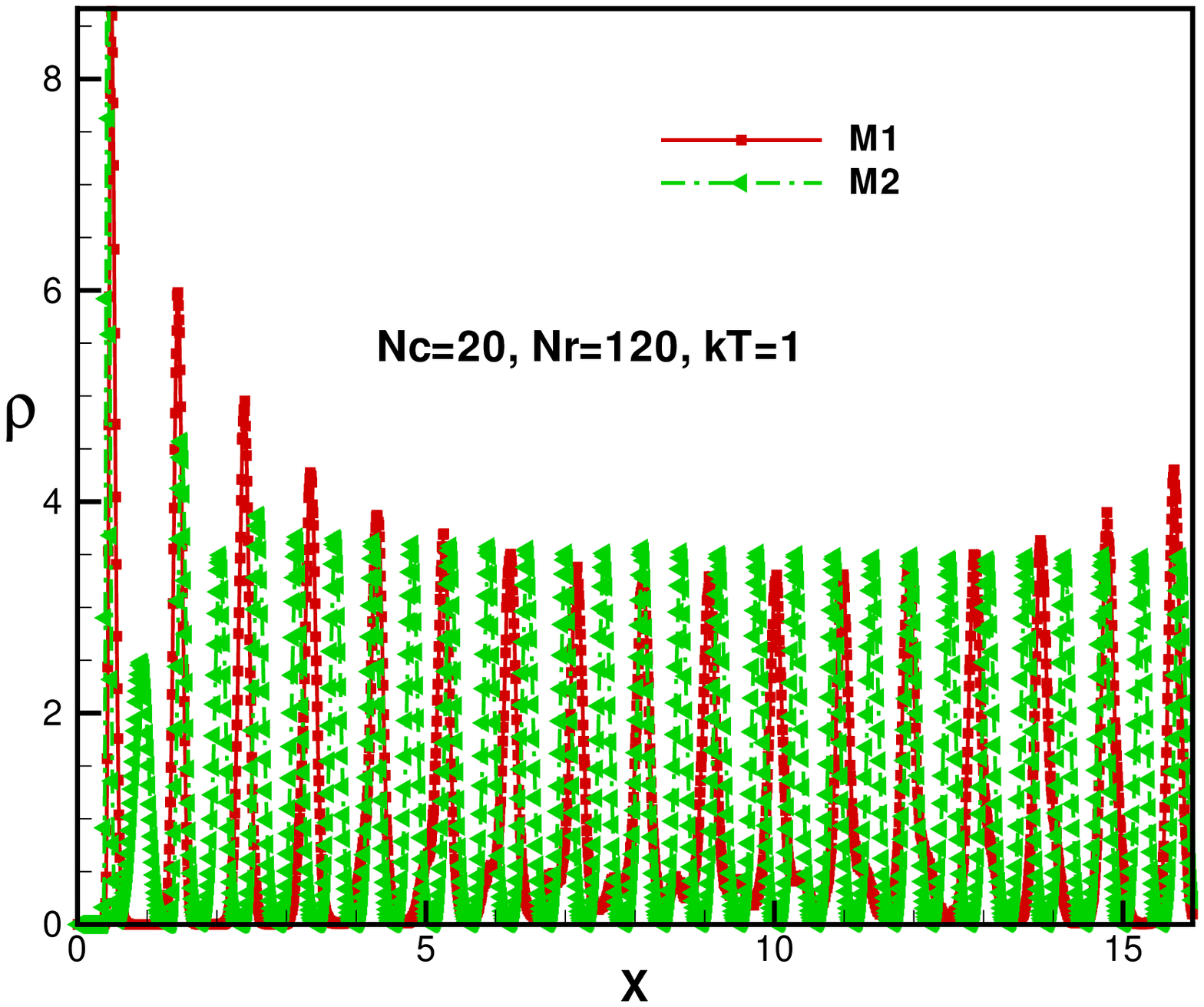}
\includegraphics[width=7.5cm]{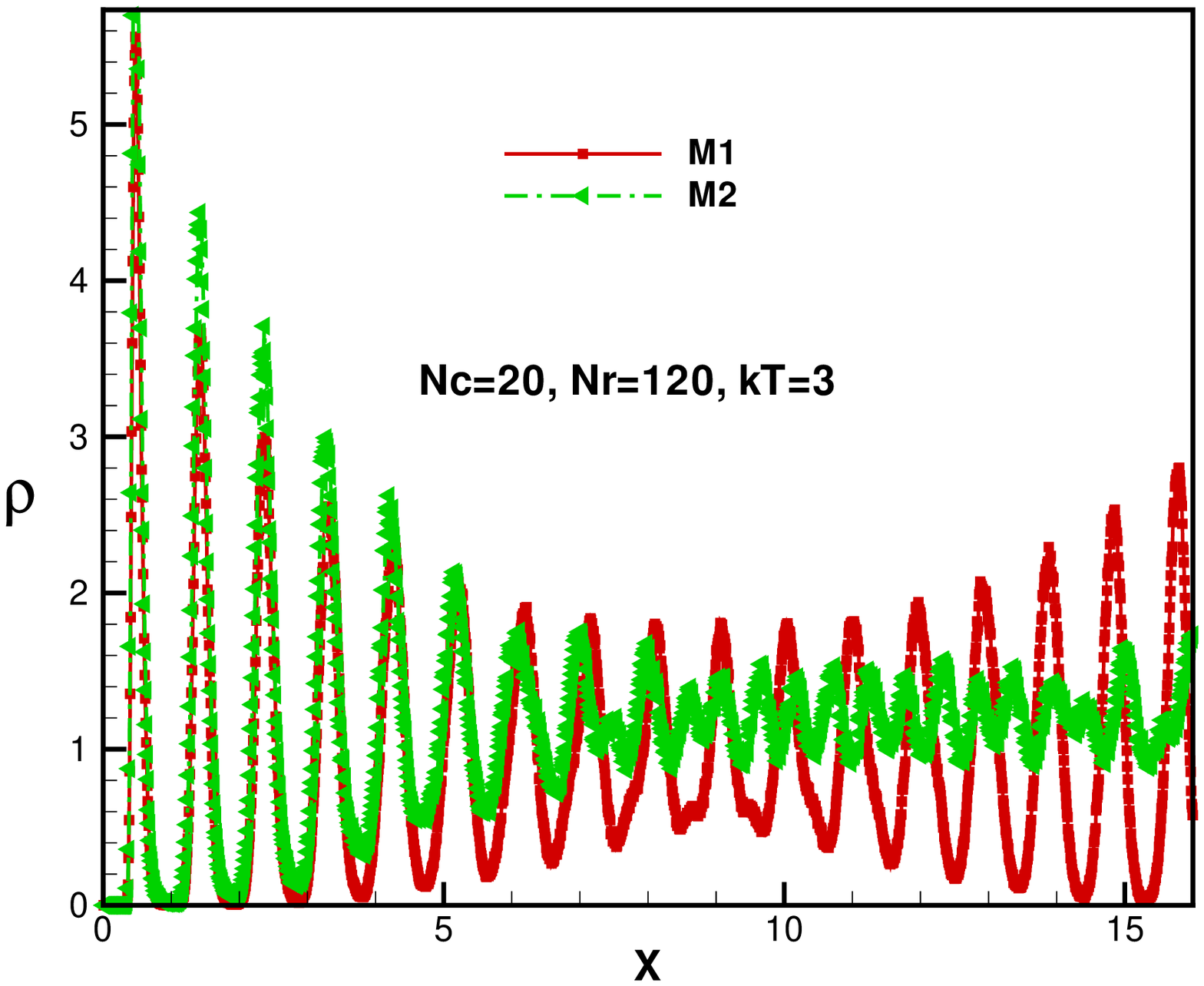}
\caption{ Density profile for a narrow channel with $N_c=20$ and $N_r=120$ at two different temperatures.
Soft wall boundary condition with both methods one and two is considered. Top: $k_BT=1$, bottom: $k_BT=3$.}
\label{fig:bz2}
\end{figure}

In figure (5) the density profiles for a temperature above the melting points for both methods
one and two as well as PBC are shown. Note in the soft wall boundary condition, the system favours to preserve its
layering structure near the walls. In the soft wall boundary condition the profiles of method one
and two are almost similar to each other. As stated earlier, in method two, the melting in the centre
is more evident.

\begin{figure}[h]
\centering
\includegraphics[width=7.5cm]{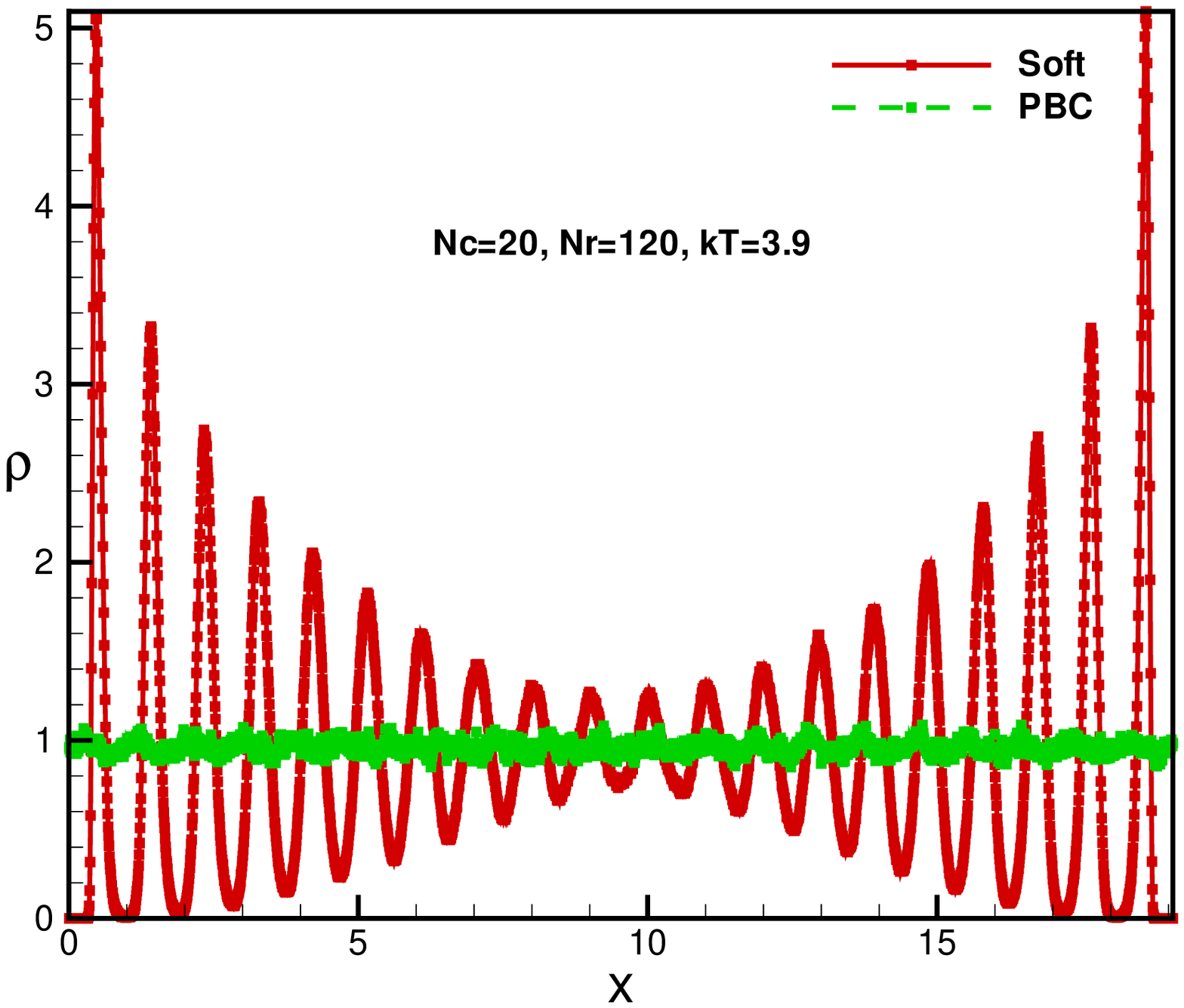}
\includegraphics[width=7.5cm]{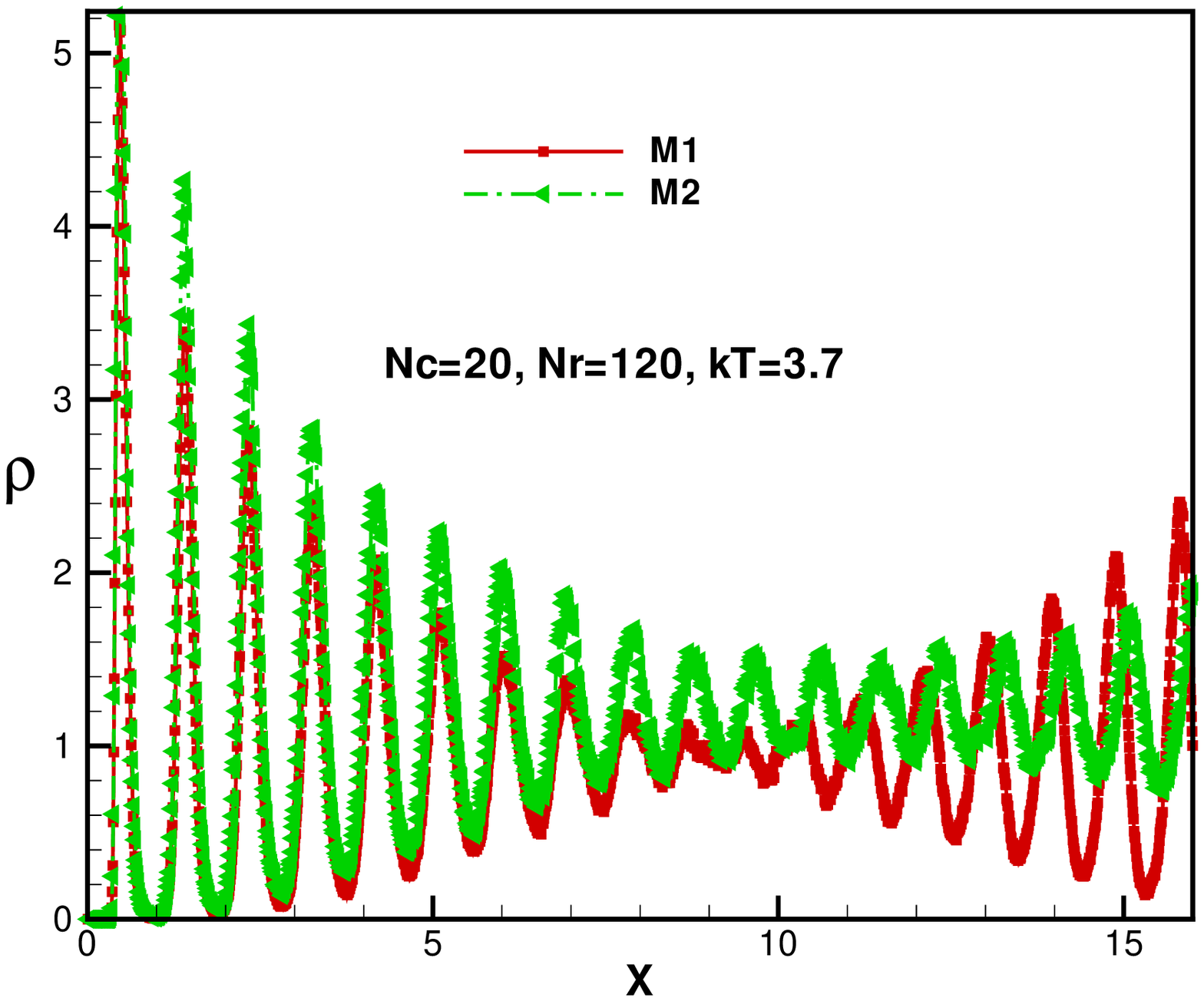}
\caption{ Top : Density profile for a narrow channel with $N_c=20$ and
$N_r=120$ at a temperature above the melting point for both
boundary conditions soft wall and periodic. bottom: comparison of methods one and two in soft wall boundary condition. } \label{fig:bz2}
\end{figure}

Next we exhibit the structure factor $S({\bf q})$ in figure (6)
for a temperature below the melting point where the colloidal
system is in the solid phase. The boundary condition is soft
walls. We recall the definition of the structure factor $S({\bf
q})$:

\begin{eqnarray}
S({\bf q})=\frac{1}{N}\sum_{l,m}\langle e^{i{\bf q}.({\bf
r}_l-{\bf r}_m)}\rangle
\end{eqnarray}

Where $\langle~\rangle$ denotes time averaging. We show this
quantity for both methods of initial setting. Note in method one
we have taken ${\bf q}=(q,0)$ whereas in method two we took ${\bf
q}=(0,q)$. The sharp peaks confirms the solid structure of the
system. The first (largest) peak is associated with the nearest
neighbour distance $a_0$.

\begin{figure}[h]
\centering
\includegraphics[width=7.5cm]{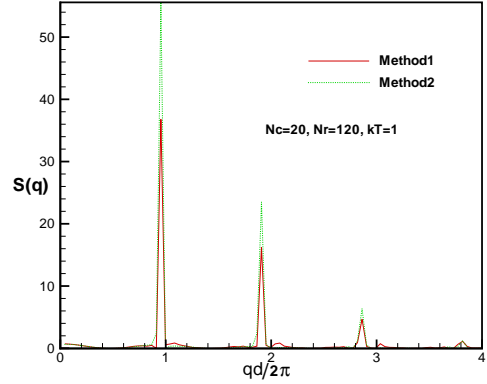}
\caption{ Structure factor $S(q)$ for a narrow channel with soft
wall boundary condition at a temperature $k_BT=1$ for initial
setting. Methods one and two give identical results. }
\label{fig:bz2}
\end{figure}

To give a quantitative description of the degree of order in the
system, we obtain the orientational order parameter $\Psi_6$. This
quantity is related to the local orientational order parameter
associated with each particle $k$.

\begin{eqnarray}
\Psi_6(k)=\frac{1}{6}\sum_{j(n.n.~ of~ k)} e^{6i\phi_{jk}}
\end{eqnarray}

Where $\phi_{jk}$ denotes the angle between a reference line
(here positive $y$ axis) and the line connecting particle $k$ to
particle $j$. Figure (7) shows the profile of orientational order
parameter squared modulus for a narrow channel at various
temperatures at $\rho=1.05$ for the soft wall system. As you can
see the modulus of $\Psi_6$ is greater near walls than in the
channel centre. Similar to density profile, the walls enhance the
degree of orientatinoal order near them. The results of Monte Carlo
simulations show quite similar behaviour \cite{binder07}. As you can
see there is notable difference between methods one and two of initial
triangular setting at high temperatures. Method one has higher orientational
order than method two which can be attributed to its higher stiffness.
The difference gets sharper when the temperature arises.

\begin{figure}[h]
\centering
\includegraphics[width=7.5cm]{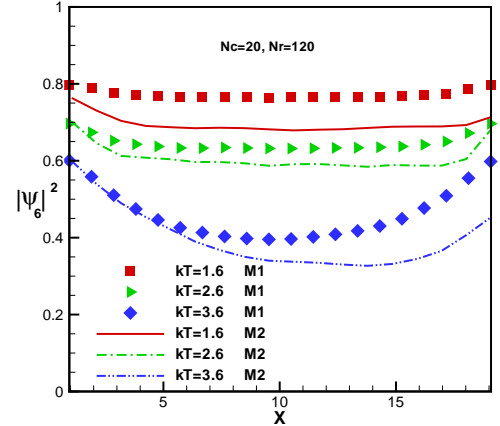}
\caption{The squared modulus of the orientational order parameter
profile $\Psi_6$ of a narrow channel with soft-wall boundary
condition at $\rho=1.05$ for various values of temperature.
Methods one and two are shown. } \label{fig:bz2}
\end{figure}

Figure (8) shows the dependence of the orientational order
parameter squared modulus at the channel centre as well as near
its walls versus $T$ at $\rho=1.05$ for both methods one and two.
These results are in qualitative agreement with MC results
\cite{binder07}. When the vicinity of the walls are considered
only a monotonous decrease with the temperature is observed. On
the contrary, when the channel centre is considered, a change in
the slope emerges which can be attributed to system melting in
the centre. By increasing the temperature, the difference between
methods one and two becomes enhanced.

\begin{figure}[h]
\centering
\includegraphics[width=7.5cm]{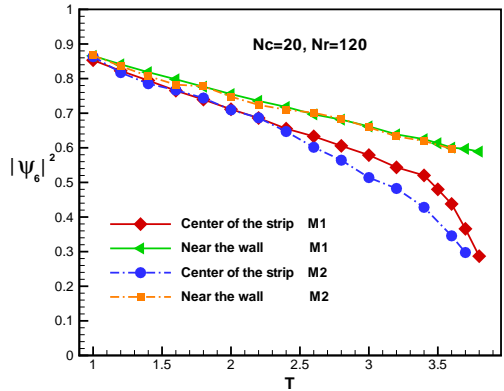}
\caption{ Temperature dependence of the orientational order
parameter squared modulus at the centre as well as near the
channel walls. Method one and two are compared to each other. }
\label{fig:bz2}
\end{figure}

\section{Elastic constants}

Apart from the study of the lattice structure, positional and
orientational order parameters and structural properties, there is
also considerable interest in the mechanical properties and in
particular elastic constants of two-dimensional crystals. The
dependence of these elastic constants on temperature (or density,
respectively) plays a crucial role in the theory of
two-dimensional melting \cite{halperin,young,binder02}. One
expects a significant effect of the symmetry of the crystal
structure. The Voigt notation has been implemented here
\cite{chaikin}. In $2d$ we have four elastic constants $C_{11},
C_{22}, C_{12}, C_{33}$. In this paper we have implemented the
method of stress fluctuation to obtain the elastic constants
\cite{squire,rafiitabar}. This method which is based on an
atomic-level description was originally introduced by Born and Huang \cite{born}.
The contribution from the particles to the elastic constants are as follows :

$$C_{\alpha\beta\gamma\delta}(i)=\frac{1}{2\Omega(i)}\sum_{j\neq i} (
\frac{\phi^{''}(r_{ij})}{r^2_{ij}} -
\frac{\phi^{'}(r_{ij})}{r^3_{ij}} )[x_\alpha(j) -
x_\alpha(i)]\times$$
$$[x_\beta(j) - x_\beta(i)][x_\gamma(j) - x_\gamma(i)][x_\delta(j)
- x_\delta(i)]$$

\begin{eqnarray} + \frac{\phi^{'}(r_{ij})}{r_{ij}}[x_\beta(j)
- x_\beta(i)][x_\gamma(j) - x_\gamma(i)]\delta_{\alpha \delta}
\end{eqnarray}

Note the term proportional to $\delta_{\alpha \delta}$ should not
be considered in the soft wall case. The contribution from a wall
is obtained from the following relation:

$$C_{\alpha\beta\gamma\delta}(i)=\frac{1}{2\Omega(i)}[\frac{1}{x_i^2}\phi_{W}^{''}(x_i)
- \frac{1}{x_i^3}\phi_{W}^{'}(x_i) ]
[x_{\alpha}^{W}-x_{\alpha}(i)]$$ \begin{eqnarray}
[x_{\beta}^{W}-x_{\beta}(i)][x_{\gamma}^{W}-x_{\gamma}(i)][x_{\lambda}^{W}-x_{\lambda}(i)]
\end{eqnarray}

In which $\phi_{W}$ is the wall potential imposed on the
particles. Also note $x_{y}^{W}=y_i$ and $x_{x}^{W}=0$ for the
left wall and $x_{x}^{W}=D$ for the right wall. Figure (9) shows
the dependence of elastic constants in a narrow channel versus
the density in the solid phase. Soft wall boundary condition is
implemented and both methods of initial triangular setting are
considered. All the components increase with increment of $\rho$.
This seems natural since the solid becomes more tough when the
density is increased. Except $C_{11}$ for which both methods give
identical results, the other three components of the elastic
tensor shows notable difference for method one and two. For
$C_{12}$ and $C_{33}$ the method two has higher value in a given
density whereas for $C_{22}$ method one value is larger than
method two.

\begin{figure}[h]
\centering
\includegraphics[width=7.5cm]{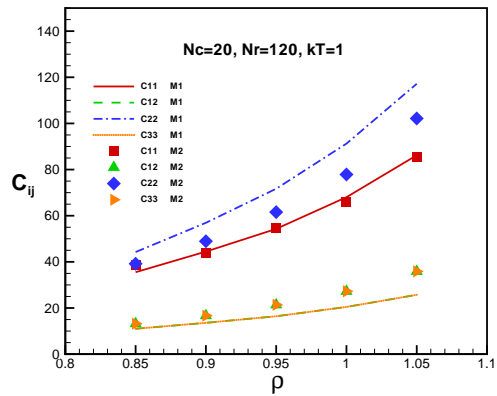}

\caption{ Dependence of the elasticity tensor components on
density in soft wall boundary condition at $k_BT=1$ for both
methods one and two. The values of the elastic constants show
substantial difference in methods one and two. } \label{fig:bz2}
\end{figure}

In order to have a further insight into the problem, we have
investigated the dependence of elastic constants at a give
density ($\rho=1.05$) on the channel width $D$. Figure (10)
exhibits the dependence of elastic constants on the number of
columns $N_c$ and the rows $N_r$.

\begin{figure}[h]
\centering
\includegraphics[width=7.5cm]{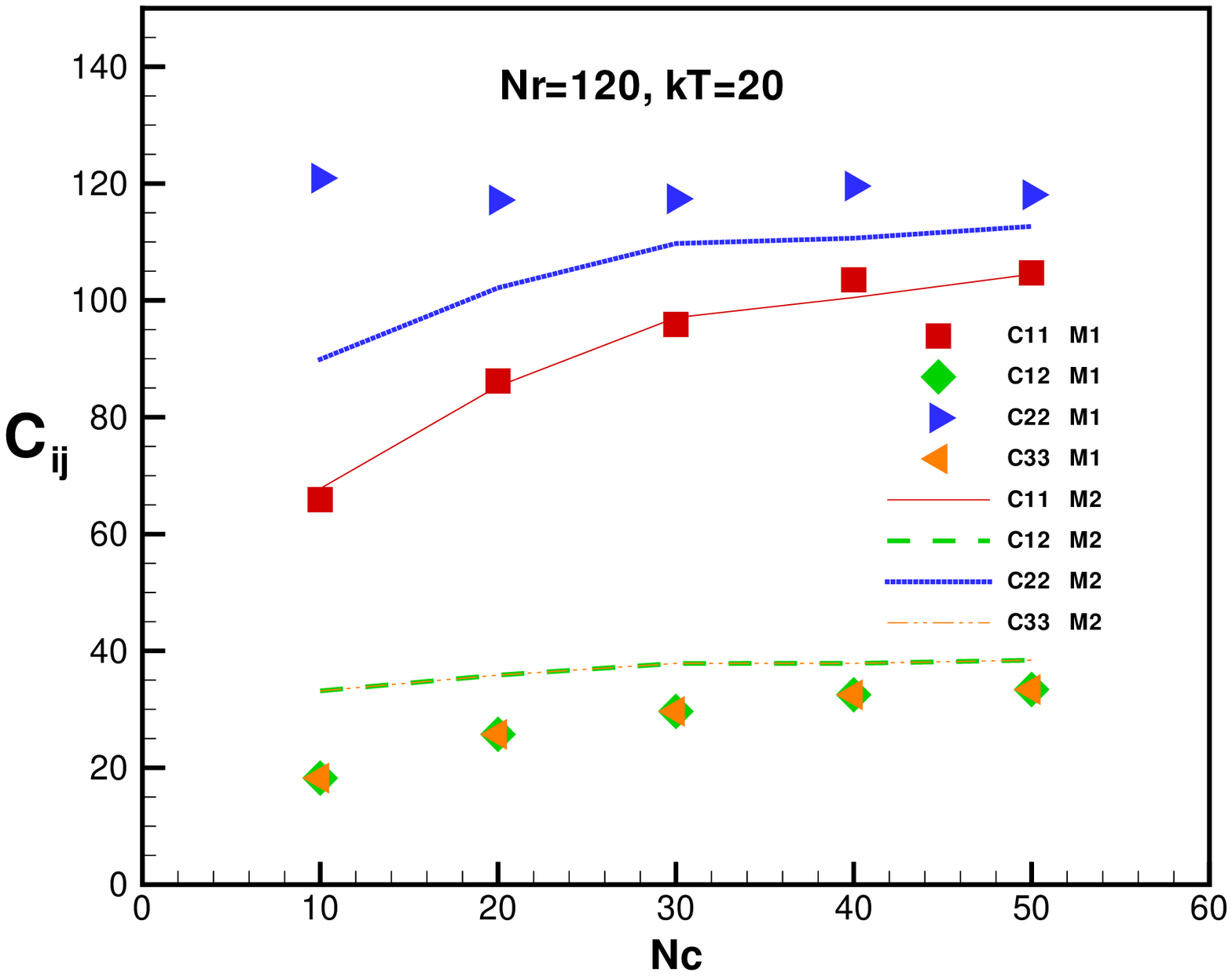}
\includegraphics[width=7.5cm]{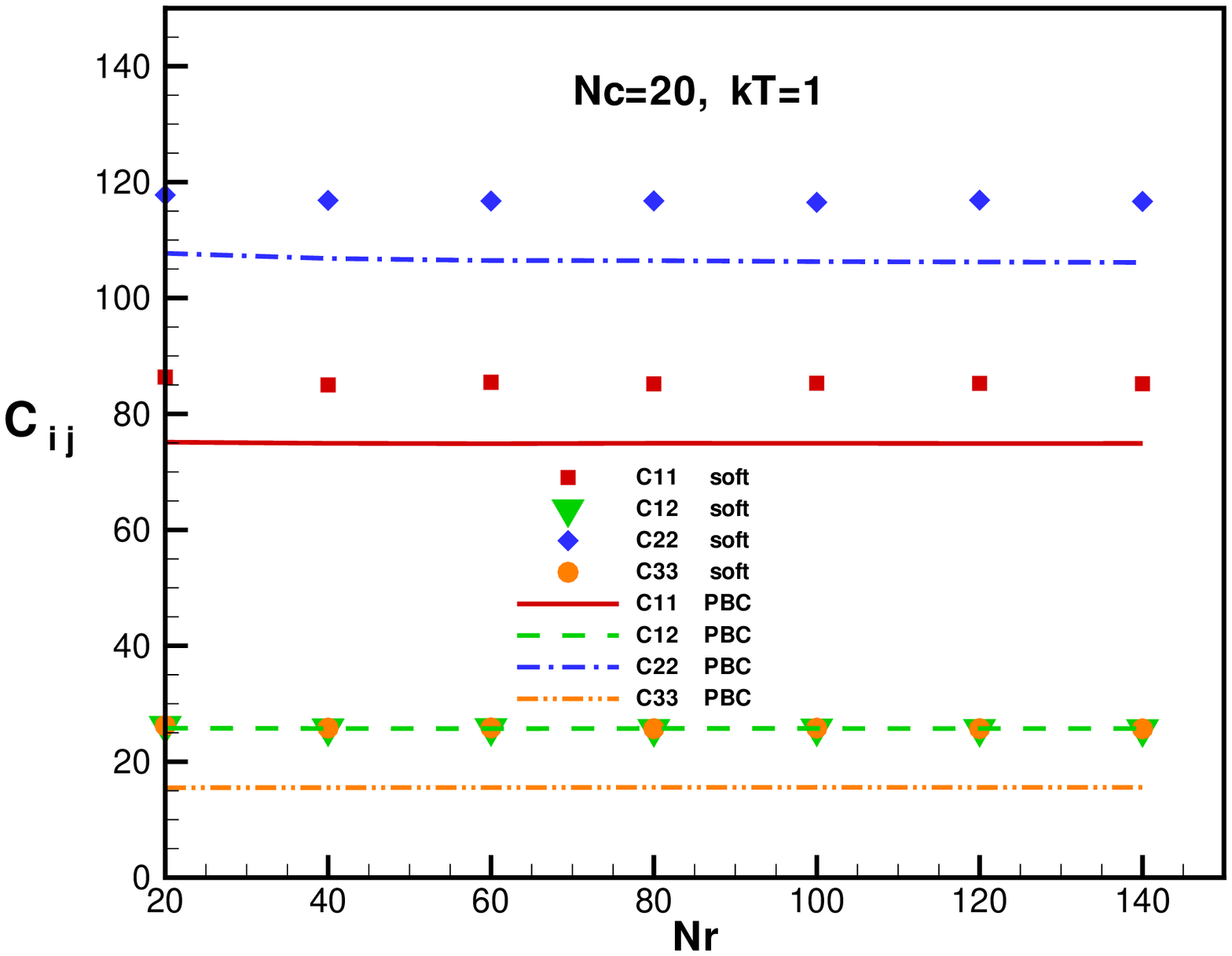}
\caption{ Top: Dependence of the elasticity tensor components on
the number of columns $N_c$ for the soft wall system at $k_BT=1$
for both methods one and two. bottom: dependence of the elastic
tensor components on the number of rows $N_r$ for both soft wall
(method one) and periodic boundary condition at $k_BT=1$. The value for the
elasticity tensor components in the PBC are smaller than the corresponding values
in the soft wall boundary condition.  }
\label{fig:bz2}
\end{figure}

The channel width $D=(N_c+1)\frac{\sqrt{3}a_0}{2}$ plays a
noticeable role. The constant $C_{11}$ is mostly affected by the
channel width. $C_{22}$ has leastly affected. After $N_c$
increases beyond $60$ (which equals to $\frac{N_r}{2}$) the
channel width $D$ plays almost no role and the values approach
the bulk ones. Our result in figure (10a) are in qualitative
agreement with those by monte Carlo simulations \cite{binder07}.
The main difference is that in our results, $C_{12}$ and $C_{33}$
coincide with each other when the system width becomes large
whereas in \cite{binder07} they do not. Notice the symmetry
$C_{12}=C_{33}$ is expected in the bulk. The values of our
elastic constants are less than those in \cite{binder07}. We
remark that the channel length in reference \cite{binder07}
($N_r=30$) is not the same as ours ($N_r=120$). Similar to
previous graph, for $C_{11}$ methods one and two give almost
identical results. Also for a given width $D$, $C_{12}$ and $C_{33}$ are larger
for two while $C_{22}$ is smaller. Molecular dynamics simulation allows us to compute the components
of the stress tensor by averaging over the system particles
trajectories. The stress tensor elements can, in principle, be
evaluated from the particles trajectories. There are two
contributions: one from the particles and other one from the
walls. The contribution from the particles to the stress tensor
component associated to particle $i$ turns out to be
\cite{rafiitabar}:

\begin{eqnarray}
\sigma_{\alpha\beta}(i)=\frac{1}{2\Omega(i)}\sum_{j \neq
i}\frac{1}{r_{ij}}\phi'(r_{ij})[x_{\alpha}(j)-x_{\alpha}(i)][x_{\beta}(j)-x_{\beta}(i)]
\end{eqnarray}

To evaluate the wall contribution, we assume each wall as a fixed
particle with infinite mass at the same height of particle $i$.
With this in mind, the contribution of the left wall which is
located at $x_{LW}=0$ becomes:

\begin{eqnarray}
\sigma^{LW}_{\alpha\beta}(i)=\frac{1}{2\Omega(i)}\phi_{LW}'(x_i)\delta_{\alpha,x}x_{\beta}(i)
\end{eqnarray}

In which $\phi_{LW}(x_i)$ is the potential energy between the left
wall and particle $i$. Similarly, the contribution from the right
wall yields to be:

\begin{eqnarray}
\sigma^{RW}_{\alpha\beta}(i)=\frac{1}{2\Omega(i)}\phi_{RW}'(D-x_i)\delta_{\alpha,x}(D-x_{\beta}(i))
\end{eqnarray}

Dependence of the stress tensor components on the density for a
temperature below melting is shown in figure (11).

\begin{figure}[h]
\centering
\includegraphics[width=7.5cm]{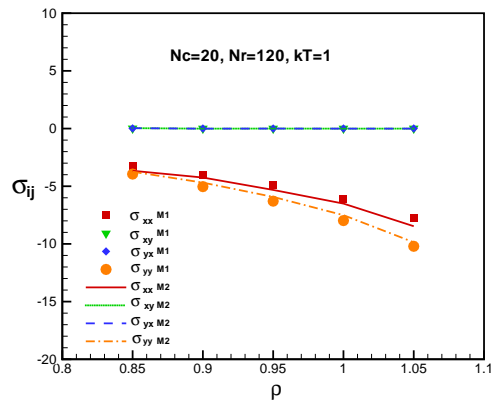}
\caption{ Dependence of the stress tensor components on the
density for a narrow channel with soft wall boundary condition in
the solid phase (method one and two are exhibited). }
\label{fig:bz2}
\end{figure}

As can be seen, the symmetry $\sigma_{xy}=\sigma_{yx}$ is
fulfilled. There is a notable dependence for the nonzero
components $\sigma_{xx}$ and $\sigma_{yy}$ on the density. By
increasing the density $\rho$, they tend to decrease. As you can see
there is not much difference between method one and two.
In figure (12) we have sketched the dependence of bulk and shear modulii
$B$ and $\mu$ on $\rho$ and $N_c$ for a narrow channel with soft wall boundary
condition for methods one and two. These quantities increase non
linearly with the density $\rho$. Again we see that beyond
$N_c=\frac{N_r}{2}$ there is almost no dependence on $N_c$. In comparison between methods one and two,
we see that the bulk modulus does not show significant change but the shear modulus does. In fact, method two gives
larger shear modulus which is expected since method two system has a smaller dgree of stiffness and hence larger shear modulus.

\begin{figure}[h]
\centering
\includegraphics[width=7.5cm]{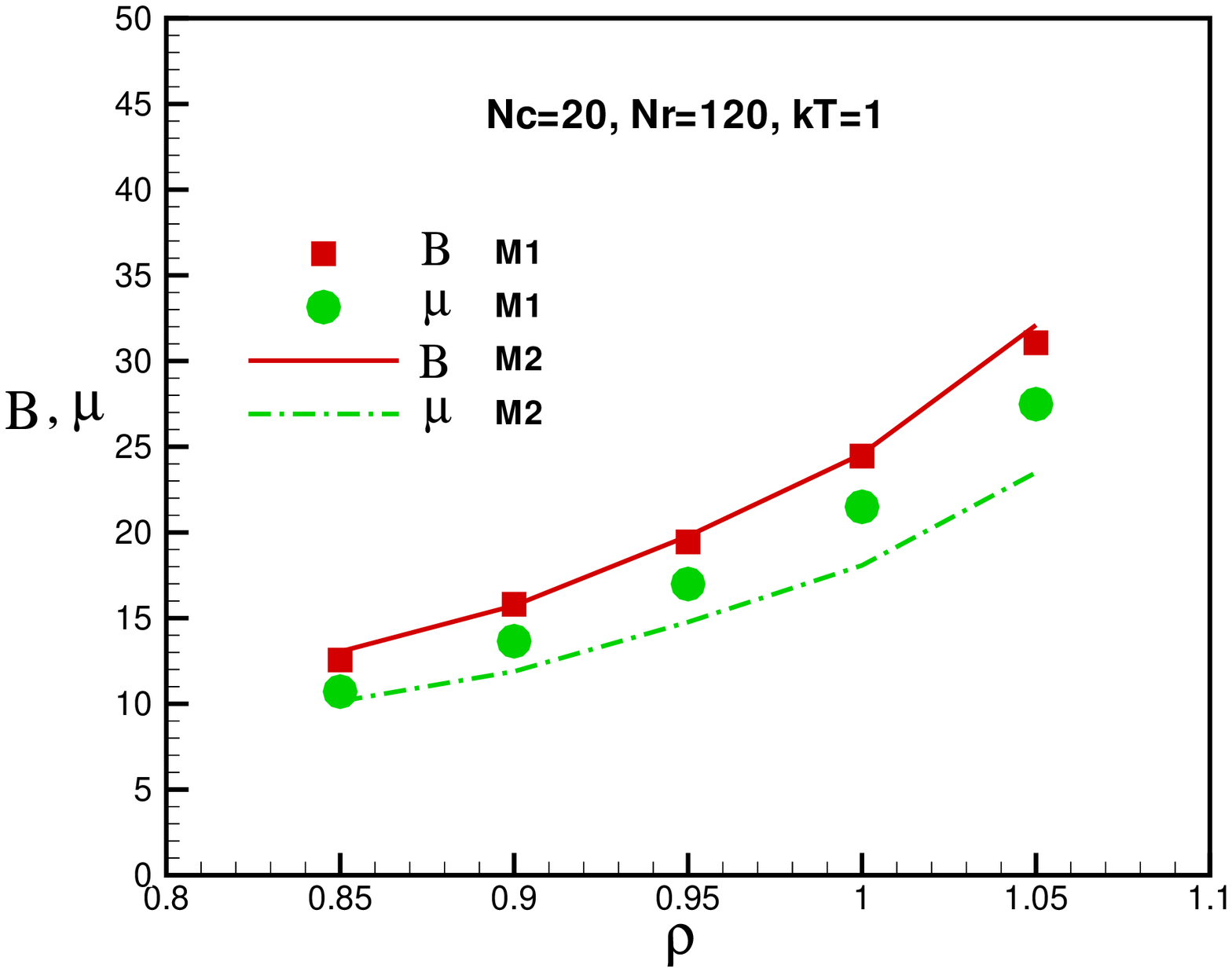}
\includegraphics[width=7.5cm]{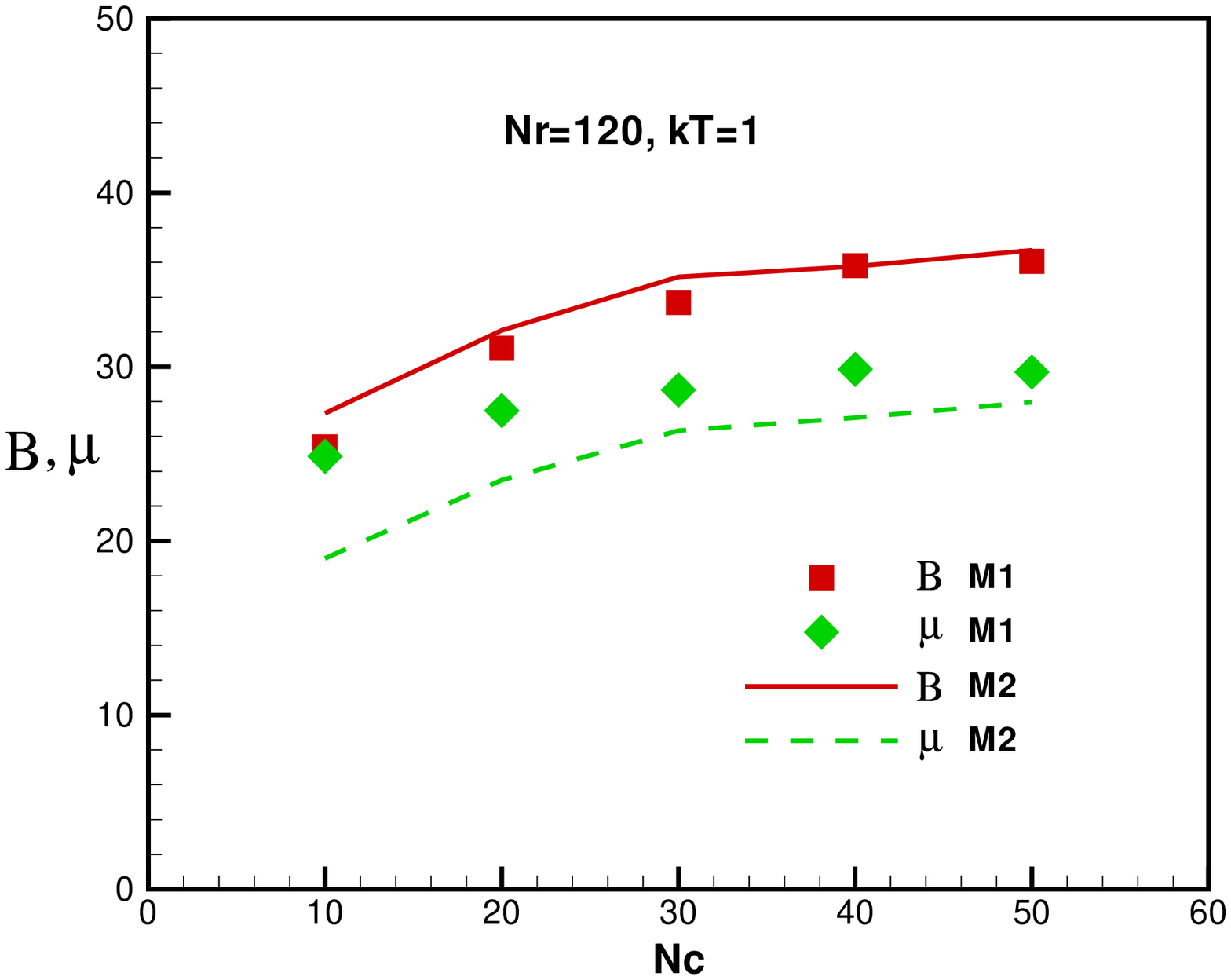}
\caption{ Top: Density dependence of the bulk and shear mudulii
$B$ and $\mu$ in the solid phase soft wall boundary condition.
Bottom: Column number dependence of the bulk and shear mudulii
$B$ and $\mu$ in the solid phase. Methods one and two are compared. } \label{fig:bz2}
\end{figure}

\section{Channel with incommensurate width to triangular lattice }

In the previous sections, the channel width $D$ was carefully
chosen such that the ideal triangular lattice structure fits into
the channel as perfectly as possible. It would be interesting to
see what happens when such a choice is not made, and $D$ does not
correspond to an integer multiple of the distance between columns
$d=\frac{\sqrt{3}a_0}{2}$ (method one). Such questions have been considered in
the literature (e.g., Refs. \cite{binder07} ) for ultra thin
strips and structures rather rich in defects were found. Here we
investigate the impact of the incommensurability on the system
characteristics. Figures (13) and (14) show the dependence of
stress and elasticity tensors components on the incommensurability
parameter $\Delta$ which is defined as $d_R=\frac{\sqrt{3}a_0}{2}
+ \Delta$. We observe the bulk and shear modulii are mostly
affected by variations of $\Delta$ whereas stress and elasticity
components are less affected. In figure (13) we see that
$\sigma_{xy}$ and $\sigma_{yx}$ do not show any significant
dependence on $\Delta$ whereas the diagonal elements
$\sigma_{xx}$ and $\sigma_{yy}$ exhibit quite noticeable
dependence on $\Delta$. In figure (14) we observe that in the
incomensurate system the elasticity tensor components decrease
when  $\Delta$ in increased.

\begin{figure}[h]
\centering
\includegraphics[width=7.5cm]{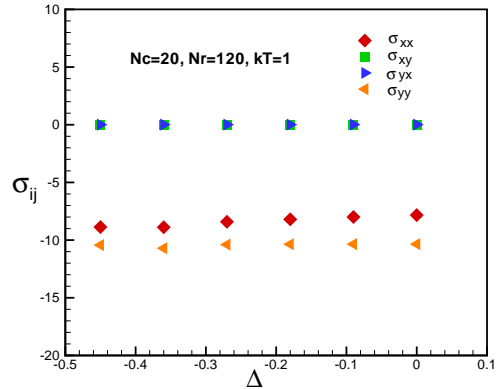}
\caption{ Dependence of the stress tensor components on the
incommensurability parameter $\Delta$. The results are for the method one
of the soft wall boundary condition. } \label{fig:bz2}
\end{figure}

\begin{figure}[h]
\centering
\includegraphics[width=7.5cm]{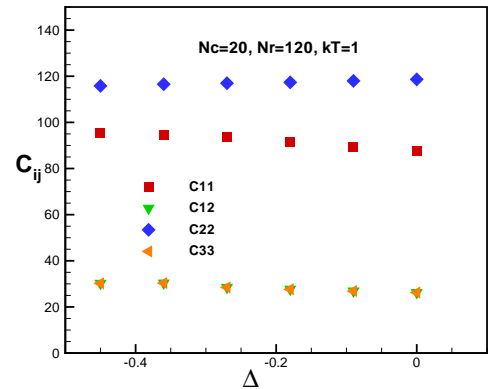}
\caption{ Dependence of the elasticity tensor components on the
incommensurability parameter $\Delta$ for the method one.} \label{fig:bz2}
\end{figure}

Eventually in figure (15) we have shown the dependence of bulk and
shear modulii on $\Delta$. Similar to elasticity tensor, by
increasing the degree of incommensurability the bulk and shear
modulii decrease. The amount of decrease is quite sharp.

\begin{figure}[h]
\centering
\includegraphics[width=7.5cm]{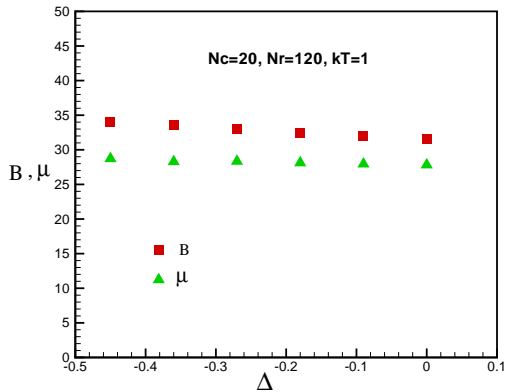}
\caption{ Dependence of $B$ and $\mu$ on the incommensurability
parameter $\Delta$ for the method one. } \label{fig:bz2}
\end{figure}

\section{Summary and conclusion}

We have used molecular dynamics simulations to study the effect
of confinement on a $2d$ crystalline solid, with triangular
structure, of point particles interacting with an inverse power
law potential proportional to $r^{-12}$ in a narrow channel. Two
methods of initial setting of particles in a triangular lattice is discussed.
The system characteristics depend sensitively on the interaction of
the two {\it walls} providing the confinement. The walls exerts
perpendicular forces on their adjacent particles. Some structural
quantities namely density profile, structure factor and
orientational order parameter are computed and their dependence
on temperature, density and other system parameters are
evaluated. It is shown that orientational order persists near the
walls even at temperatures where the system in the bulk is in the
fluid state. Moreover, the dependence of elastic constants,
stress tensor elements, shear and bulk modulii on density as well
as the channel width is discussed and is shown they increase with
raising the density. The effect of varying the channel width is
explored and it is found that in general the bulk and shear
modulii increase with increasing the channel width until the
width becomes comparable to the system length. Furthermore, the
effect of incommensurability of the channel with the triangular
lattice structure is discussed. It is shown that incommensurability
notably affects the system properties. We compare our findings to
those obtained by Monte Carlo simulations in \cite{binder07} and
also to the periodic boundary condition along the channel.
.

\section{Acknowledgement}

We are highly indebted to Professor Surajit Sengupta for his
valuable comments and fruitful discussions. We wish to express
our gratitude to Prof. Hashem Rafii Tabar and Dr Abbas Montazeri
for their useful helps and enlightening discussions. M.E.F is
thankful to No'rooz Khan for his valuable discussions.
.

\end{document}